\documentclass{moriond}

\usepackage[T1]{fontenc}
\usepackage[english]{babel}
\usepackage{lmodern}

\usepackage{amsmath,amsfonts,amsthm} 

\def\Journal#1#2#3#4{{#1} {\bf #2}, #3 (#4)}

\def\NCA{\em Nuovo Cimento}

\def\NPB{{\em Nucl. Phys.} B}
\def\PLB{{\em Phys. Lett.}  B}
\def\PRL{\em Phys. Rev. Lett.}
\def\PRD{{\em Phys. Rev.} D}

\newcommand{\paren}[1]{\left( #1 \right)}
\newcommand{\ele}{\mathcal{L}}

\newcommand{\Photo}{}

\begin{document}
\vspace*{4cm}
\title{LINEAR {\Large\boldmath$\sigma$} MODEL FOR THE GOLDSTONE H -- EXPERIMENTAL TESTS}  

\author{S. SAA}

\address{Departamento de F\'isica Te\'orica and Instituto de F\'isica Te\'orica, IFT-UAM/CSIC,
Universidad Aut\'onoma de Madrid, Cantoblanco, 28049, Madrid, Spain}

\maketitle\abstracts{
In order to explore a possible dynamical nature for the Higgs field (such as its being a 
pseudo-Goldstone boson) we develop a renormalizable Lagrangian based on the minimal 
$SO(5)$ linear $\sigma$-model 
with the symmetry softly broken to $SO(4)$, including gauge bosons and 
fermions. We then present the phenomenological implications and constraints from 
precision observables and the impact on present and future LHC data.
}

%

\section{Motivation}\label{sec:motiv}

The LHC data in 2012 showed the existence of a spin zero particle of mass
around $125$~GeV~\cite{aa,ch}. This has been identified with the boson of the
Brout-Englert-Higgs mechanism~\cite{eb,h1,h2}
(for simplicity, from here on just called the ``Higgs boson''),
responsible for the mass of the Standard Model (SM) particles. 

However, theoretical considerations like the so-called
electroweak (EW) hierarchy problem
pose questions about the naturalness of an elementary scalar. 
In particular, about the lightness of the Higgs boson
with respect to any  higher new scale that couples to it.
The simpler way to sort this out is to try to protect its mass through a symmetry,
like in supersymmetry, composite Higgs, little Higgs...
The Higgs boson is thus an excellent window to 
look into the dynamics of the spontaneous electroweak symmetry breaking (EWSB).

In this work we focus on a framework in which the Higgs is a pseudo-Goldstone boson.
This idea was suggested by the fact that the only other elemental 
scalars already existing in the SM
are the Goldstone bosons of the EWSB.
In particular, our model is inspired by the composite Higgs framework~\footnote{
Also in theories such as ``little Higgs'' or models based on extra dimensions
 the Higgs is considered to have a Goldstone boson origin.},
first proposed in a seminal paper~\cite{kg}, where
all Goldstone bosons originate from the spontaneous breaking of a global $SU(5)$
group into $SO(5)$ at some high scale~$\Lambda$; such that $\Lambda\leq4\pi f$,
where $f$ corresponds to the pion scale $f_\pi$,
in analogy to chiral symmetry breaking in QCD.
Recent implementations  seek to produce just the minimal set that provides
the Higgs and the three longitudinal components of the EW gauge bosons,
which is achieved by an $SO(5)\to SO(4)$ breaking scheme~\cite{acp}.

In the absence of an explicit breaking of the global symmetry the Higgs boson is massless,
as the other three Goldstone bosons generated. Therefore, some
additional explicit breaking of $SO(5)$ is needed. 
This is provided through the coupling of the $SO(5)$ sector~\footnote{
Besides the scalar multiplet, 
the $SO(5)$ sector contains also some vectorial heavy fermions which will mix
directly to the SM ones. For brevity, we will omit here the details of these;
which can be consulted in the paper~\cite{fgk}.}
to the SM fermions and gauge bosons and results in an effective potential 
whose minimum breaks spontaneously the EW symmetry at scale $v$,
which is determined by the Fermi decay constant. 
This scale is in general different from $f$, being the difference between
them a known source of fine-tuning in this type of theories.
The breaking also provides the Higgs particle with a mass
so it becomes a pseudo-Goldstone boson.

Most of the literature on composite Higgs uses an effective non-linear approach
to study the models~\cite{crp,cm,mss,pan,car}.
Instead, we study here a renormalizable model, a particular UV completion of those,
which includes a new scalar, $\sigma$, singlet under the EW gauge group. 
This extra scalar gets a mass due to the spontaneous breaking of $SO(5)$.
By varying it one can sweep from the linear, weakly coupled regime 
(light $\sigma$ particle) to the non-linear one ($m_\sigma\to\infty$),
where one should fall onto the standard effective approach.
Another advantage of this model is that it might be
considered as a renormalizable UV completion of some deeper dynamics; such
as the so-called linear $\sigma$ model for QCD at low energies~\cite{gml};
but it can be also  regarded as a renormalizable model made out of elementary fields.

%

\section{The $SO(5)/SO(4)$ scalar sector}\label{sect:scalar}

We have defined~\cite{fgk} a complete Lagrangian composed of pure
gauge, scalar and fermionic parts and their interactions.
However, here we
will just focus on the scalar sector; which instead of being
the usual SM Higgs sector is now substituted by a Higgs-$\sigma$ sector.
Thus we will consider the five scalars in the fundamental of $SO(5)$:
\begin{equation}
\phi= \left(\pi_1,\pi_2,\pi_3,h,\sigma\right)^T 
	\quad\stackrel{u.g.}{\to}\quad \left(0,0,0,h,\sigma\right)^T\, ~,
\end{equation}
where $h$ corresponds to the Higgs particle and $\sigma$ is the extra scalar 
aforementioned.
For simplicity, from now on, we will work in the unitary gauge (u.g.); where
the three components, $\pi_i$, associated with the longitudinal components of
the EW gauge bosons are set to zero.

The potential 
\begin{equation}
V(h,\sigma) = \lambda\paren{h^2+\sigma^2-f^2}^2+\alpha f^3\,\sigma-\beta f^2\,h^2\,
 \label{eq:Laghs}\,
\end{equation}
contains one term ($\lambda$) parametrizing the spontaneous breaking
and two terms ($\alpha$ and $\beta$) which break $SO(5)$ explicitly, while preserving 
the $SO(4)$ symmetry. The inclusion of these terms can be argued through
the computation of the one-loop Coleman-Weinberg effective potential;
since they are needed as counterterms to absorb the divergences generated
(see Appendix in the full paper~\cite{fgk}).
Terms like these have also been used in a previous attempt in this direction~\cite{bb},
where phenomenology is discussed in a simpler fermionic setup.

Due to the explicit breaking, both $\sigma$ and $h$ acquire a non-trivial vev 
and, as a consequence, they mix in the light and heavy mass eigenstates, which 
will be parametrized by an angle $\gamma$.

From the kinetic scalar Lagrangian, which contains the interaction with
the SM gauge bosons through the covariant derivative, it is easy to show
that the vev of $h$ should correspond to the EW scale in order to yield 
the observed $W$ and $Z$ masses
\begin{equation}
\ele_{\text{s, kin}} = \frac{1}{2} (D_\mu\phi)^T (D^\mu\phi)\,
\qquad\longrightarrow\qquad \langle h \rangle = v= 246~\text{GeV}~.
\label{eq:Lkin}
\end{equation}
\begin{figure}[h]
\centering
\includegraphics[width=0.5\textwidth,keepaspectratio]{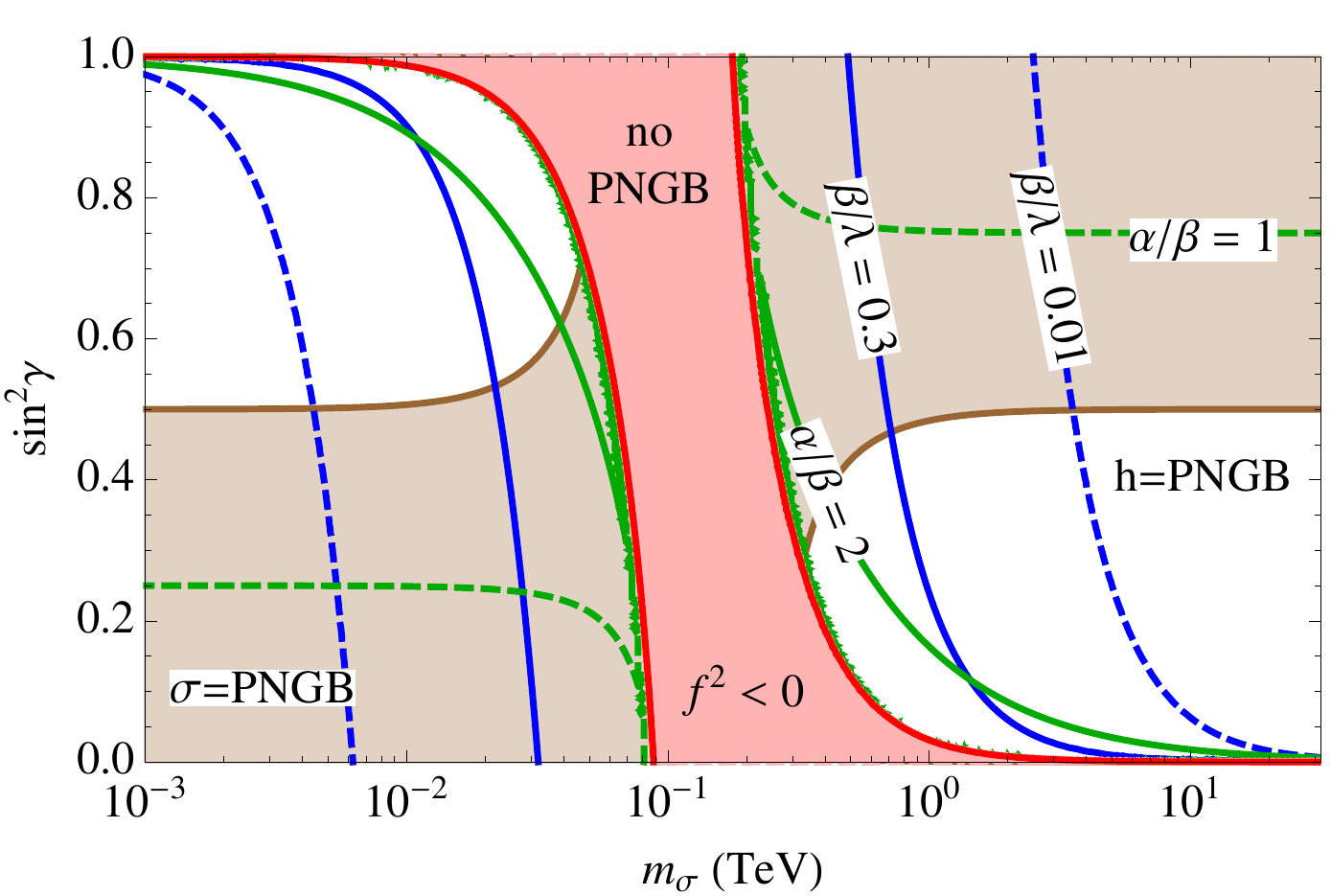}
\caption[]{Parameter space for the scalar mixing and $\sigma$ mass for the
observed Higgs mass and vev. The brown region corresponds to 
$\langle \sigma \rangle\equiv v_\sigma<v$;
while in the white region, $v<v_\sigma$. The pseudo-Goldstone boson in each region
corresponds to that with the smaller vev. 
In the red region, $f^2<0$, and thus the spontaneous
symmetry breaking is lost. The relative importance of the explicit breaking
is shown through the $\beta/\lambda$ and $\alpha/\beta$ curves.}
\label{fig:Vplot}
\end{figure}
In order to understand the parameter space of the model, we show in Fig.~\ref{fig:Vplot}
the scalar mixing versus the mass of $\sigma$ for the known Higgs mass and vev.
We find that the Higgs being a pseudo-Goldstone boson philosophy only
holds in the white region on the bottom right corner of the plot,
where $h$ has the smaller vev and mass.
There is another region, on the top left corner, 
where $h$ is also a pseudo-Goldstone boson;  despite being heavier.
This is because in this area the allowed mixing is large,
thus inverting the $h$ and $\sigma$ character
inside the mass eigenstates. 
Despite its phenomenological interest, due to naturalness reasons,
we will focus in the $m_\sigma>m_h$ white area in what follows.

%

\section{Phenomenology}

\begin{figure}[h]
\centering
\includegraphics[width=0.39\textwidth,keepaspectratio]{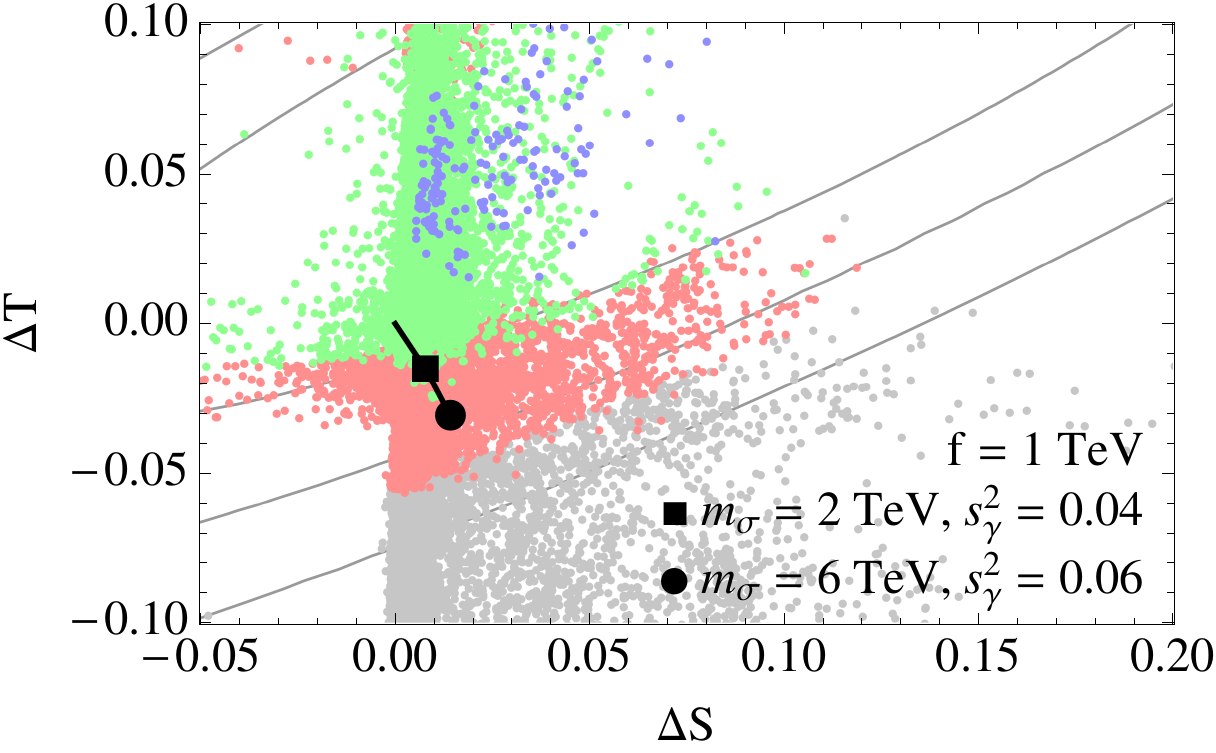}\hfill
\includegraphics[width=0.3\textwidth,keepaspectratio]{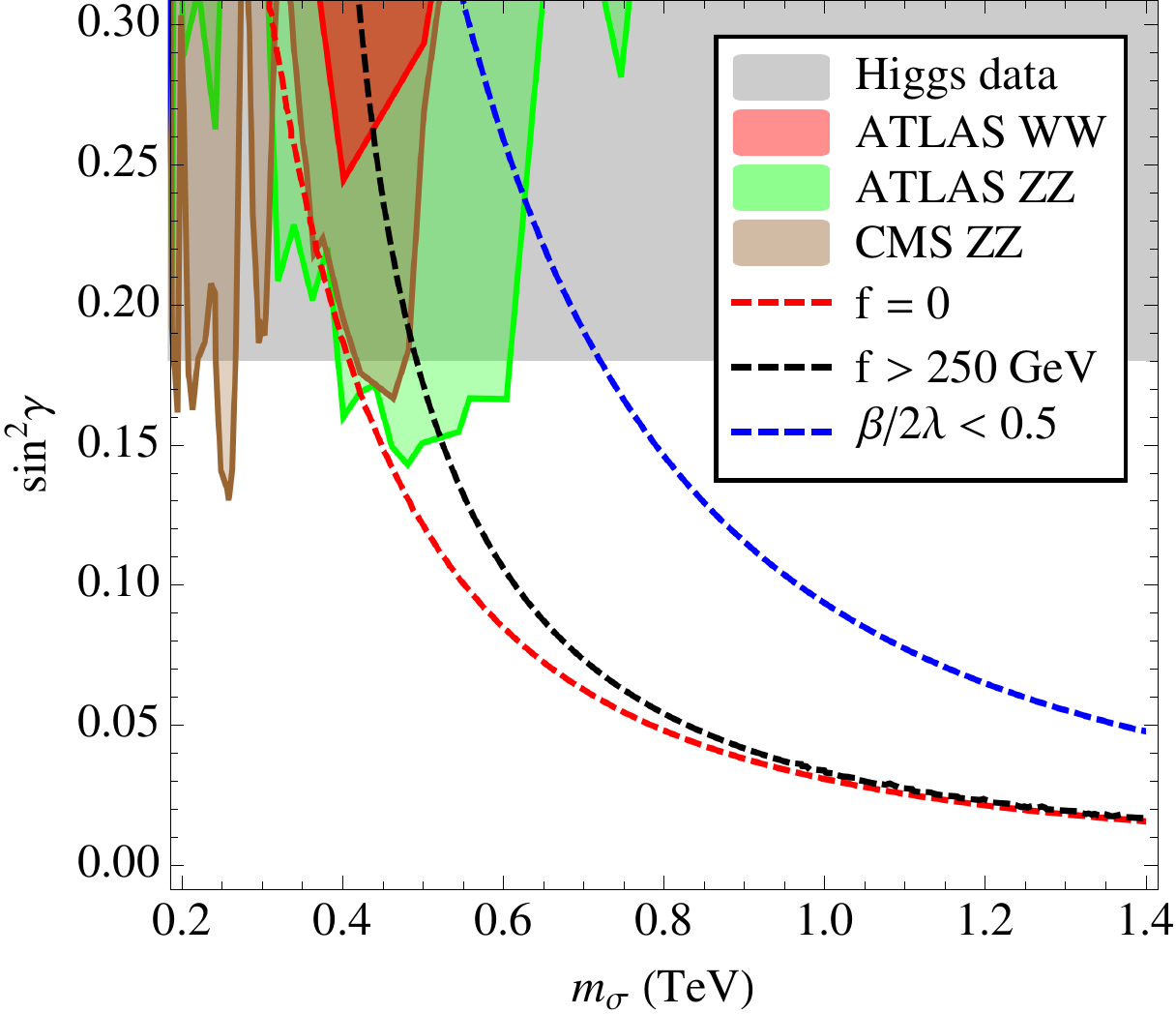}
\includegraphics[width=0.3\textwidth,keepaspectratio]{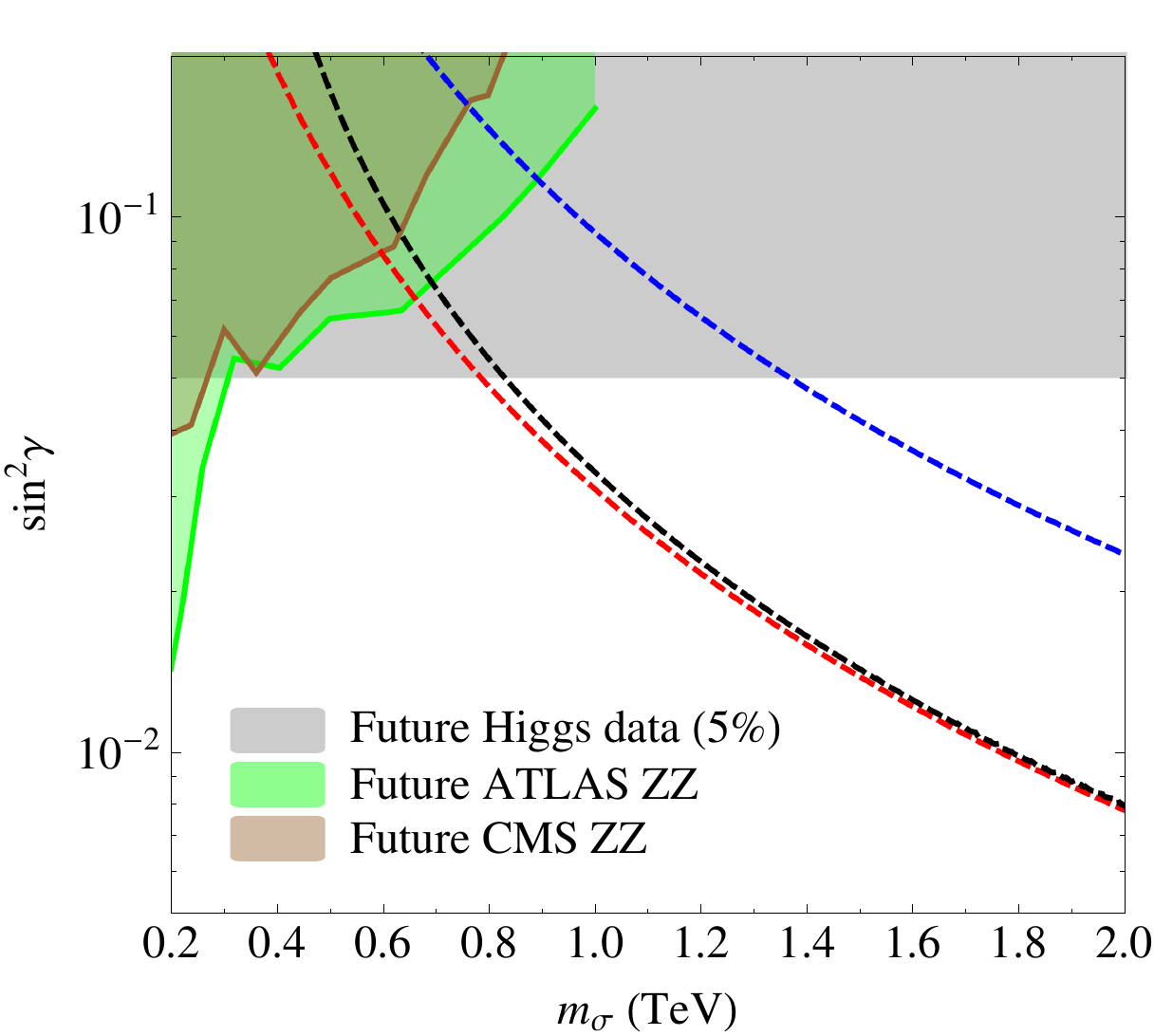}
\caption[]{Left: coloured points stand for the contribution from exotic fermions alone;
blue, green, red and grey 
represent $1\sigma$, $2\sigma$, $3\sigma$ and $>4\sigma$ deviations from the SM,
respectively; from a global fit to S, T and the modification to $Zb_Lb_L$,
and black dots are the contribution of the $\sigma$ scalar.
Centre: bounds on scalar parameters from current LHC data  
(fundamentally diboson searches~\cite{atlasWW,cmsWW,atlasZZ,cmsZZ}).
Right: future prospects for the $14$~TeV LHC run
with an integrated luminosity of 3 $ab^{-1}$~\cite{ho}, 
assuming a 5\% precision on Higgs couplings. 
The production of the $\sigma$ particle is assumed to be dominated by gluon
fusion, due to the large gluon pdfs. This is somewhat model-dependent,
depending on the coupling between the $\sigma$ and the exotic 
fermions.}
\label{fig:pheno}
\end{figure}

\paragraph{Precision tests}\mbox{}\\[0.5em]
We have computed the contributions of scalars and exotic 
fermions to precision parameters $\Delta S$ and $\Delta T$~\cite{pt}.
In brief, as shown in the left plot in Fig.~\ref{fig:pheno}, 
the fermionic contribution is typically quite spread in $\Delta T$; while the 
deviations from the SM due to the $\sigma$ particle push
$\Delta S$ and $\Delta T$ towards positive and negative values, respectively.
Thus, in general, including a lighter $\sigma$ in the model
will help alleviate the tension with precision parameters.
%
\paragraph{The \boldmath$\sigma$ particle and LHC data} \mbox{}\\[0.5em]
In the centre and right plots in Fig.~\ref{fig:pheno} 
we show  the bounds from different LHC searches for
heavy scalars translated
onto the parameter space of our model.
We have determined the excluded grey area from the modification
to the couplings of $h$ to fermions and gauge bosons~\cite{ac}.

In contrast to the precision tests, where a lighter $\sigma$ is preferred,
in order to consider the Higgs as a pseudo-Goldstone boson,
the parameters allowed in the plots
need to be to the right of the black curve. This sets a lower bound of around
$m_\sigma>550$ GeV from current data, which can be pushed above
$900$ GeV in the future.
This is a consequence of the underlying symmetry of the model;
while in analyses for a generic scalar singlet that mixes with $h$
the allowed region would be the whole white area.

\paragraph{Can the \boldmath$\sigma$ be the 750~GeV diphoton excess?}%
\mbox{}\\[0.5em]
The $750$ GeV diphoton excess observed by ATLAS and CMS~\cite{ac}
could be explained by a zero-spin resonance such as the $\sigma$ scalar.
However, there are reasons to consider this somewhat unnatural.
First, both the production through gluon fusion and the decay into photons
are loop induced and too small to account for the signal unless the mixing 
is made extremely suppressed in order to prevent
a large decay into $WW$ and $ZZ$.
A tiny mixing angle leads to a very fine-tuned
$\alpha$ against $\beta$ parameter and, moreover, for $750$ GeV, we would fall
onto the red area in figure~\ref{fig:Vplot}, where no spontaneous symmetry breaking
takes place.
Another way out would be to provide a larger amount of heavy fermions to increase
the multiplicities in the loops.

%
\section{Conclusions}
We have analysed in depth a linear framework for the
Higgs as pseudo-Goldstone boson.
The model contains an extra scalar, $\sigma$, whose mass acts as the
ultraviolet scale.
A light $\sigma$ particle is found to help decrease part of the tension with precision tests;
while LHC data together with naturalness considerations impose a lower bound on
its mass.

\section*{Acknowledgments}

My work is supported through
the grant BES-2013-066480 of the Spanish MICINN
within the research project FPA2012-31880
and by the Spanish MINECO's ``Centro de Excelencia Severo Ochoa" 
Programme under the grant SEV-2012-0249.
Special thanks to the other collaborators in Madrid (Bel\'en Gavela and
Pedro Machado) and Padova (Ferruccio Feruglio, Kirill Kanshin and Stefano Rigolin).
I finally want to thank the organisers of the Moriond conference for
making the meeting such a successful and enjoyable event.

\end{document}